\documentclass[twocolumn,english,journal]{IEEEtran}
\usepackage{babel}
\usepackage{calc}
\usepackage{amssymb}
\usepackage{graphicx}
\usepackage[unicode=true,
 bookmarks=true,bookmarksnumbered=true,bookmarksopen=true,bookmarksopenlevel=1,
 breaklinks=false,pdfborder={0 0 0},pdfborderstyle={},backref=false,colorlinks=false]
 {hyperref}
\hypersetup{pdftitle={Your Title},
 pdfauthor={Your Name},
 pdfpagelayout=OneColumn, pdfnewwindow=true, pdfstartview=XYZ, plainpages=false}

\makeatletter

\providecommand{\tabularnewline}{\\}

 \let\oldforeign@language\foreign@language
 \DeclareRobustCommand{\foreign@language}[1]{%
   \lowercase{\oldforeign@language{#1}}}

\usepackage[caption=false,font=footnotesize]{subfig}

\makeatother

\begin{document}

\title{Environmental Modeling of Silicon Dangling Bond QCA Wires}

\author{Dan Brox, Post-Doctorate research completed in the Department of Electrical and Computer Engineering, University of British Columbia, Vancouver, BC, Canada.} 

\maketitle
\begin{abstract}
Interactions of quantum cellular automata (QCA) circuits with their
environment induce transitions in their quantum states that can cause
errors in computation. The nature of these interactions depend on
the specific physical implementation of the circuit. In the case of
silicon dangling bond QCA, one channel of environmental interaction
is between dangling bond cells and longitudinal phonons in the silicon
lattice. In the presence of this environmental interaction, short
4 cell wire simulations show reliable operation at liquid nitrogen
temperature, however simulation and theoretical arguments suggest
long wires operated from thermal equilibrium are susceptible to exponential
decay of cell polarization along their length regardless of clock
zoning. Quantum annealing is suggested as a technique for bringing
QCA wires into their ground state prior to information transmission
to circumvent this problem. 
\end{abstract}

\begin{IEEEkeywords}
quantum, cellular automata, dangling bond, phonon
\end{IEEEkeywords}

\IEEEpeerreviewmaketitle{}

\section{Introduction}

\IEEEPARstart{C}{}omputing schemes based on electronic quantum cellular
automata (QCA) have been proposed as a route to performing classical
computation higher speeds and lower power consumption than achievable
with transistor-based architectures {[}1{]}. These schemes encode
classical information via the electronic states of atomic-scale ``cells''
for which various physical implementations have been suggested {[}2-4{]}.
QCA cells typically interact with each other electrostatically to
alter each other's states, and so by carefully arranging cells in
networks that constrain these interactions, circuits are created that
digitally process the states of input cells to produce states of output
cells, thereby performing computation. 

\noindent An important feature of QCA computations is that it is desirable
to maintain QCA circuits in their ground state throughout computation
so that the final states of output cells are correctly determined
by the states of input cells. A process of gradually modulating the
tunneling barriers of cells while input cell polarizations are switched
has been proposed as a means of maintaining the ground state. In this
process, known as adiabatic switching {[}1{]}, a circuit beginning
in its ground state is guaranteed to be found in its ground state
at the end of computation with high probability if switching occurs
slowly enough, according to the adiabatic theorem of quantum mechanics
{[}5{]}.

\noindent One difficulty with the adiabatic clocking scheme is that
although it can maintain a closed QCA circuit in its ground state,
the adiabatic theorem does not account for interactions of the circuit
with its environment that may cause transitions to excited states.
This problem persists even when a circuit is held in a static configuration
in which equilibrium thermal effects are easily computed. For instance,
consider a QCA wire with Hamiltonian:

\begin{equation}
H_{wire}=-P_{in}\frac{E_{k}}{2}\sigma_{z}^{1}-\frac{E_{k}}{2}\sum_{i=1}^{N-1}\sigma_{z}^{i}\sigma_{z}^{i+1},
\end{equation}

\noindent where each cell $i$ is in a $|0\rangle$ or $|1\rangle$
$\sigma_{z}^{i}$ eigenstate, $E_{k}$ is the energetic cost of having
adjacent cells in different states, and $P_{in}$ is the input polarization
that affects the state of cell 1 and thereby influences the states
of all cells along the wire. If we assume that the input polarization
is $+1$, the ground state of the wire is $|1,1,1,....1,1\rangle$
(\textit{N} 1's). In contrast, excited states of the wire contain
a nonzero number of kinks, and we can expect that the last cell of
the wire is in the $|0\rangle$ or $|1\rangle$ state depending on
whether the total number of kinks is odd or even. If the wire is in
thermal equilibrium with its surroundings at temperature T, then the
Boltzmann factor associated with a kink of energy $E_{k}$ is $p=e^{-\frac{E_{k}}{k_{B}T}}$,
and the probabilities of the $M^{th}$ cell in the wire occupying
the $|0\rangle$ or $|1\rangle$ are calculated to be:

\begin{eqnarray}
Prob(|0\rangle) & = & \frac{\left(1+p\right)^{M}-\left(1-p\right)^{M}}{2\left(1+p\right)^{M}},\\
Prob(|1\rangle) & = & \frac{\left(1+p\right)^{M}+\left(1-p\right)^{M}}{2\left(1+p\right)^{M}}.
\end{eqnarray}

\noindent This means that the expected polarization of the $M^{th}$
(output) cell is:

\begin{equation}
\frac{Prob(|1\rangle)-Prob(|0\rangle)}{Prob(|1\rangle)+Prob(|0\rangle)}=\frac{\left(1-p\right)^{M}}{\left(1+p\right)^{M}},
\end{equation}

\noindent which decays to 0 as $M\rightarrow\infty$. This indicates
that for sufficiently long wires, in the absence of clocking (i.e.
time independent Hamiltonian), thermal effects can completely destroy
the information we expect to see propagated by the wire.

\noindent This same thermal issue also poses problems for any sufficiently
large QCA circuit in a static state, so it is interesting to ask if
the dynamics of QCA computation can be employed to prevent thermal
degradation. To this end, it has been suggested that if a QCA circuit
is divided into sufficiently small individually clocked subcircuits
or ``zones'', then each subcircuit can perform computation with
high fidelity so that appropriate circuit architecture can be used
to solve the problem {[}1{]}. However, it is not clear that this zoning
hypothesis is true, and little simulation work has been done to verify
this is true for specific circuit-environment interactions and zoning
schemes. Furthermore, better understanding of this issue is critical
to deciding how large QCA circuits can be made before other classical
computing elements must be included into their architecture {[}6{]}
to make useful devices. Therefore, in this paper we investigate the
effect of zoning and electron-phonon interactions on a silicon dangling
bond QCA wire to ascertain how effective zoning is in preventing thermal
degradation in this particular case.

\section{QCA Wire Zoning}

\noindent To introduce the notion, we consider a simple silicon dangling
bond QCA circuit which we'll take to be a straight wire. Each cell
in this wire consists of two silicon dangling bonds and a single electron
which is localized at either site during latching and shared between
sites during cell relaxation. A schematic of an 8 cell wire partitioned
into four zones in shown in Fig. 1.

\begin{center}
\includegraphics[scale=1.0]{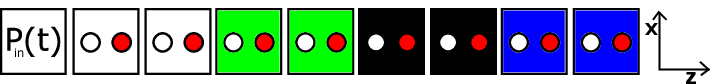}
\par\end{center}

\begin{flushleft}
{\small{}Fig. 1 Schematic of an 8 cell silicon dangling bond QCA wire
partitioned into four zones.}
\par\end{flushleft}{\small \par}

\noindent Since we are interested in adiabatically clocking the cells
to transmit information along the wire, the relevant time dependent
Hamiltonian for the wire is:

\begin{eqnarray}
H_{wire}(t) & = & -P_{in}(t)\frac{E_{k}}{2}\sigma_{z}^{1}+\frac{E_{k}}{2}\sum_{i=1}^{N}\gamma^{i}(t)\sigma_{x}^{i}\\
 &  & -\frac{E_{k}}{2}\sum_{i=1}^{N-1}\sigma_{z}^{i}\sigma_{z}^{i+1},\nonumber 
\end{eqnarray}

\noindent where the input polarization is now time dependent and tunneling
modulators $\gamma^{i}(t)$ are employed to latch and delatch the
states of the cells. For specificity, we choose a kink energy of $E_{k}=0.05$
eV and maximum tunneling modulation $max|\gamma^{i}(t)|=1.0$. These
values are representative of a QCA wire composed of silicon dangling
bond qubits {[}4{]}. It is important to note that the adiabatic clocking
assumed in equation (5) has not been physically implemented in any
silicon dangling bond circuit to date, and technical challenges to
modulating the intercell tunneling as necessary are presented by the
atomic dimensions of the cells. Nevertheless, silicon dangling bond
technology is still in early stages of development, so a standard
form of the QCA wire Hamiltonian has been adopted here for ease of
comparison with other theoretical work and to suggest specifications
on the modulation in anticipation of a technique for physically implementing
it.

\noindent Simulation of different zoning schemes are made by choosing
different input polarization functions and tunneling modulation functions.
For instance, the Fig. 2 shows an 8 cell wire divided into 4 clocking
zones, and four (normalized) clocking signals Clock 1 ($\gamma^{1}(t)=\gamma^{2}(t)$),
Clock 2 ($\gamma^{3}(t)=\gamma^{4}(t)$), Clock 3 ($\gamma^{5}(t)=\gamma^{6}(t)$),
and Clock 4 ($\gamma^{7}(t)=\gamma^{8}(t)$).

\begin{center}
\includegraphics[scale=1.0]{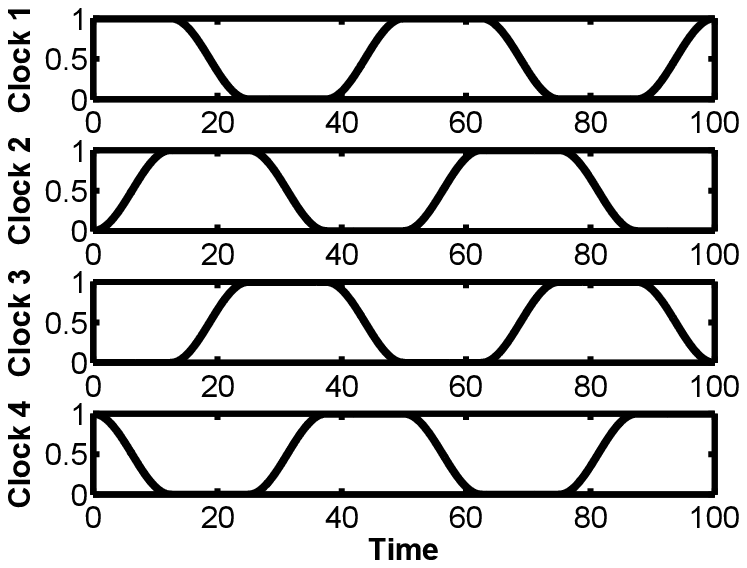}
\par\end{center}

\begin{flushleft}
{\small{}Fig. 2 Four phase clocking of a QCA wire.}
\par\end{flushleft}{\small \par}

\noindent The clocking signals are shown over 2 complete periods in
time units of $\hbar/E_{k}$. When a clock signal is low, the corresponding
cell is latched in a fixed $\sigma_{z}$ eigenstate, and when it is
high the corresponding cell is in a relaxed unlatched state where
it can be influenced by the state of its neighbors. This is the reason
for applying 4 clock phases in succession to iteratively latch and
unlatch data along the wire and pipeline information flow.

\noindent For the purposes of investigating the interplay between
wire zoning and phonon interaction, we can consider simple information
transmission processes in which the input polarization is fixed at
1 throughout. To do this, we begin with the each wire cell in its
relaxed state $\gamma^{i}(0)=\gamma_{max}$, and switch the cells
from their relaxed state to hold states in various ways, examining
the expected polarization of each of the cells throughout. For instance,
in the case of a four cell wire, we could use 1 clocking zone and
clock all the cells to the hold state simultaneously, or use 2 clocking
zones to clock cells 1 and 2 to a hold state and lag the clocking
of cells 3 and 4 to the hold state by a quarter phase as in the 4
phase clocking scheme. We could also use 4 zones, separately clocking
each of the zones with a quarter lag phase between the switching to
the hold state of each. This clocking scheme is illustrated in Fig.
3. Note that in this particular scheme, once a particular zone is
latched, it remains latched for the rest of the process.

\begin{center}
\includegraphics[scale=1.0]{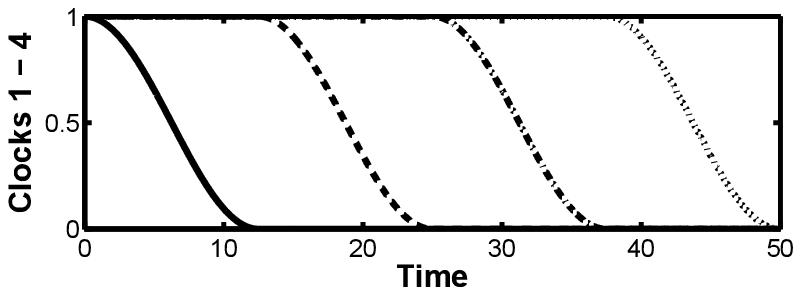}
\par\end{center}

\begin{flushleft}
{\small{}Fig. 3 Sequential latching of a QCA wire divided into 4 clocking
zones. }
\par\end{flushleft}{\small \par}

\noindent Regardless of the clocking scheme employed, the $t=0$ initial
state of the wire for each process is the same, and in this state
the expected polarization of the cells approaches 0 as we move further
from the driver cell as indicated in Table 1.

\begin{center}
{\small{}TABLE I: Initial polarizations of cells in a relaxed four
cell wire with input polarization 1.}
\par\end{center}{\small \par}

\begin{center}
\begin{tabular}{|c|c|c|c|c|}
\hline 
 & Cell 1 & Cell 2 & Cell 3 & Cell 4\tabularnewline
\hline 
\hline 
Initial Polarization & 0.8276 & 0.7221 & 0.6078 & 0.4472\tabularnewline
\hline 
\end{tabular}
\par\end{center}

\noindent Therefore, it is necessary to latch the cells of the wire
into hold states so that their polarizations are able to follow the
input polarization and convey information along the wire.

\section{Phonon Interaction}

\noindent To better understand environmental effects on silicon dangling
bond QCA circuits, we must first specify a particular interaction,
which we take here to be the interaction between the cells fabricated
on a silicon crystal surface and phonons propagating through the silicon
lattice. For simplicity, we'll only consider the primary influence
of longitudinal phonons in the silicon lattice on the dangling bond
QCA wire. Unlike transversely polarized phonons, longitudinal phonons
generate charge density variations by compressing and stretching the
lattice in their direction of motion, thereby interacting electrostatically
with electrons. 

\noindent \smallskip

\noindent To model this, we first write the Hamiltonian of the QCA
wire interacting with the longitudinal phonons in the silicon lattice
{[}7{]}:

\begin{equation}
H_{total}=H_{wire}+H_{int}+H_{ph},
\end{equation}

\noindent where:

\begin{eqnarray}
H_{wire}(t) & = & -P_{in}(t)\frac{E_{k}}{2}\sigma_{z}^{1}+\frac{E_{k}}{2}\sum_{i=1}^{N}\gamma^{i}(t)\sigma_{x}^{i}\\
 &  & -\frac{E_{k}}{2}\sum_{i=1}^{N-1}\sigma_{z}^{i}\sigma_{z}^{i+1},\nonumber 
\end{eqnarray}

\begin{equation}
H_{ph}=\sum_{\vec{q}}\hbar\omega_{q}\left(a_{\vec{q}}^{\dagger}a_{\vec{q}}+\frac{1}{2}\right),
\end{equation}

\noindent and the energy of interaction with electrons, known as the
deformation potential energy, is given by (c.c denotes complex conjugate): 

\begin{eqnarray}
H_{int} & = & \int d^{3}rZ_{DP}\psi^{\dagger}(r)\psi(r)\\
 &  & \cdot\left(\sum_{\vec{q}}iq\sqrt{\frac{\hslash}{2\Omega V\omega_{q}}}a_{\vec{q}}e^{i\vec{q}\cdot r}+c.c.\right).\nonumber 
\end{eqnarray}

\noindent In these equations, $\vec{q}$ indexes all possible wave
vectors of the longitudinal phonon modes, $a_{\vec{q}}^{\dagger}$
and $a_{\vec{q}}$ are the creation and annihilation operators of
a phonon in a particular mode, $\hbar\omega_{q}$ is the energy of
the phonon mode $\vec{q}$ dependent on the wave vector length $q=\parallel\overrightarrow{q}\parallel$,
$Z_{DP}\approx3.3$ eV is the deformation potential in silicon, $\Omega$
is the density of silicon, $V$ is the total volume, and:

\begin{equation}
\psi(r)=\sum_{i,\sigma}\phi_{i\sigma}(r)c_{i\sigma},
\end{equation}

\begin{equation}
\psi^{\dagger}(r)=\sum_{i,\sigma}\phi_{i\sigma}^{*}(r)c_{i\sigma}^{\dagger},
\end{equation}

\begin{equation}
[c_{i\sigma},c_{j\sigma}^{\dagger}]=\delta_{ij},
\end{equation}

\noindent are fields that create and annihilate electons at point
\textbf{r}. The indices $(i,\sigma)$ index single electron states
localized at QCA cell $i$ with symmetry $\sigma$. $\phi_{i\sigma}(r)$
is the wave function corresponding to this electron. The operator
$n(r)=\psi^{\dagger}(r)\psi(r)$ is the local electron density at
point \textbf{r}. 

\noindent \smallskip

\noindent The next task is to use the Hamiltonian for the QCA wire-phonon
system to derive an approximate equation for the quantum evolution
of the QCA wire alone. In so doing, we must appreciate that the quantum
state of the wire is entangled with the phonon state of its environment,
and so if we wish to describe its quantum state independently of the
environment we should do so with a density matrix. More precisely,
we should describe the state of the entire QCA wire-phonon system
with a density matrix $\rho_{system}$ whose evolution is described
by the Von Neumann equation:

\begin{equation}
\frac{d}{dt}\rho_{system}=-\frac{i}{\hslash}[H_{total},\rho_{system}],
\end{equation}

\noindent and use some procedure of averaging over environmental effects
to obtain an approximate equation of evolution for the density matrix
of the wire $\rho_{wire}$ alone. For our purposes, we take this approximate
equation to be a Lindblad master equation for the density matrix of
the QCA wire {[}8{]}:

\begin{eqnarray}
\frac{d}{dt}\rho_{wire} & = & -\frac{i}{\hslash}[H_{wire},\rho_{wire}]\\
 &  & +\sum_{r,s}\frac{1}{2}W_{rs}\left(2P_{rs}\rho_{wire}P_{sr}-[P_{sr}P_{rs},\rho_{wire}]\right),\nonumber 
\end{eqnarray}

\noindent where $r$ and $s$ index different (instantaneous) eigenstates
of the QCA wire, $P_{rs}$ is the state transition operator:

\begin{equation}
P_{rs}=|r\rangle\langle s|,
\end{equation}

\noindent and the constants $W_{rs}$ are the environmentally induced
transition rates from state $|s\rangle$ to $|r\rangle$ {[}9-10{]}:

\begin{eqnarray}
W_{rs} & = & \sum_{\vec{q}}\frac{\pi Z_{DP}^{2}q^{2}}{\Omega V\omega_{q}}|\langle r|\beta(\vec{q})|s\rangle|^{2}n_{ph}\left(E_{q},T\right)\\
 &  & \cdot\delta\left(E_{q}-(E_{r}-E_{s})\right),\thinspace\thinspace E_{r}>E_{s},\nonumber 
\end{eqnarray}

\begin{eqnarray}
W_{rs} & = & \sum_{\vec{q}}\frac{\pi Z_{DP}^{2}q^{2}}{\Omega V\omega_{q}}|\langle r|\beta^{\dagger}(\vec{q})|s\rangle|^{2}\\
 &  & \cdot\left(n_{ph}\left(E_{q},T\right)+1\right)\delta\left(E_{q}-(E_{s}-E_{r})\right),\thinspace\thinspace E_{r}<E_{s}.\nonumber 
\end{eqnarray}

\noindent In these formulae, $Z_{DP}$ is the deformation potential
energy ($3.3$ eV in silicon), $\Omega$ is the density of silicon
$2.33\times10^{3}$ kg$\cdot$m\textsuperscript{-3}, $V$ is the
total volume of silicon, $T$ is the temperature, $c_{s}=9.0\times10^{3}$
m$\cdot$s\textsuperscript{-1} is the speed of sound in silicon,
and $E_{q}=\hbar\omega_{q}=\hbar c_{s}q$ is the energy difference
between wire states $|s\rangle$ and $|r\rangle$. The phonon thermal
occupation number $n_{ph}\left(E_{q},T\right)$ is:

\begin{equation}
n_{ph}\left(E_{q},T\right)=\left(e^{\hslash c_{s}q/k_{B}T}-1\right)^{-1},
\end{equation}

\noindent and the QCA cell transition operator $\beta(\vec{q})$ is
defined as:

\begin{equation}
\beta(\vec{q})=\sum_{i}\sigma_{z}^{i}\langle S_{i}|e^{i\vec{q}\cdot r}|A_{i}\rangle+\sigma_{x}^{i}\langle L_{i}|e^{i\vec{q}\cdot r}|R_{i}\rangle,
\end{equation}

\noindent where the matrix elements are between different electronic
states of the QCA cells defined as:

\begin{equation}
|L_{i}\rangle\leftrightarrow\frac{1}{\sqrt{\pi a_{B}^{3}}}e^{-r_{Li}/a_{B}},\thinspace\thinspace r_{Li}=\left\Vert \vec{r}_{i}+\frac{1}{2}D\overrightarrow{k}\right\Vert ,
\end{equation}

\begin{equation}
|R_{i}\rangle\longleftrightarrow\frac{1}{\sqrt{\pi a_{B}^{3}}}e^{-r_{Ri}/a_{B}},\thinspace r_{Ri}=\left\Vert \vec{r}_{i}-\frac{1}{2}D\overrightarrow{k}\right\Vert ,
\end{equation}

\begin{equation}
|S_{i}\rangle=\frac{1}{\sqrt{2}}\left(|L_{i}\rangle+|R_{i}\rangle\right),
\end{equation}

\begin{equation}
|A_{i}\rangle=\frac{1}{\sqrt{2}}\left(|L_{i}\rangle-|R_{i}\rangle\right).
\end{equation}

\noindent In these equations,$\vec{k}$ is the unit vector in \textit{z}
direction, $L$ is the distance between cells in the wire, and $r_{i}=iL\vec{k}$
is the position of the $i^{th}$ cell. 

\noindent As an example, we can compute the transition rates for a
single cell wire at $T=300$ K. For a single cell, the kink energy
and its dependence on $L$ does not factor into the system Hamiltonian,
so it suffices to adopt an intercell atomic distance and tunneling
(i.e. relaxation) energy to determine the rates. Therefore, choosing
$D=7.68\times10^{-10}$ m and a tunneling energy of $0.05$ eV, we
obtain: 

\begin{eqnarray}
W_{10} & = & 3.60\times10^{12}\thinspace s^{-1},\\
W_{01} & = & 2.46\times10^{13}\thinspace s^{-1}.
\end{eqnarray}

\noindent In the requisite calculations, a renormalized Bohr radius
of $a_{B}=2.79\times10^{-10}$ m is used to ensure the value of the
expected tunneling determined by the electron wave functions is in
fact $0.05$ eV. In more general calculations, the Bohr radii $a_{B}$
of each cell is similarly renormalized at each time step according
to the tunnelings $\gamma^{i}(t)$, influencing the $\beta$ operators
and induced transition rates betwen wire states $W_{rs}$ and $W_{sr}$.
Note that the $\beta$ operator can be thought of as a perturbation
of $H_{total}$ induced by stretching and compressing of the silicon
lattice, and this is appropriately thought of as a modulation of intercell
tunneling, not intracell electrostatic coupling {[}11{]}. For this
reason, although we consider 2-atom QCA cells here for simplicity,
we can expect phonon induced transitions to have an effect of similar
magnitude on more standard 4-atom QCA cells. Also note that for sufficiently
long wires, the Born approximation used in deriving the Lindblad master
equation is no longer valid (see Appendix A), and in such cases the
Lindblad equation must be regarded as a hypothetical model of the
system-environment interaction rather than a rigorous approximation
of the true system evolution.

\section{Zoning and Transition Rates}

\noindent We can now examine the effect of phonon interaction on the
expected polarization wire cells at the end of latching. To do this
we need to assert the initial state of the wire, which we take to
be the ground state of a four cell wire with fixed input polarization
1 and all cells relaxed. As an initial study, we only enable transitions
between the ground and first four excited states of the wire ($W_{sr},\thinspace W_{rs}=0$
unless $r=0$ and $s$ $\in\left\{ 1,2,3,4\right\} $), and perform
switching at $T=300$ K, $k_{B}T=0.026$ eV over a total adiabatic
clocking time of $20\hbar/E_{k}=1.31\cdot10^{-12}$ s\textsuperscript{-1}.
This does not take into account the full complexity of the environmental
interaction bringing the wire to a state of thermal equilibrium, but
does account for the primary excitation channels of the ground state
into excited states. Note that the reverse excited-to-ground state
transition rates are related to the ground-to-excited state rates
by Boltzmann factors:

\begin{equation}
W_{rs}=W_{sr}\cdot e^{\left(E_{s}-E_{r}\right)/k_{B}T},
\end{equation}

\noindent to ensure detailed balance between the state populations
at thermal equilibrium.

\noindent Under our assumptions, we can use the Lindblad equation
to compute and plot the cell output polarizations at the end of each
clocking process shown in Fig. 4.

\begin{center}
\includegraphics[scale=1.0]{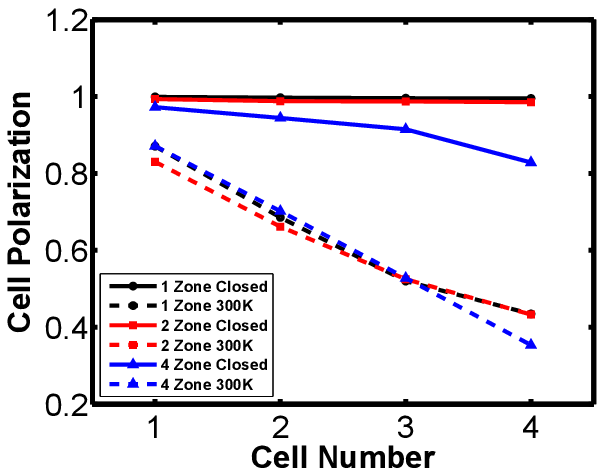}
\par\end{center}

\begin{flushleft}
{\small{}Fig. 4 Cell polarizations of four cell wire at the completion
of different clocking processes run at T=300 K.}
\par\end{flushleft}{\small \par}

\noindent In Fig. 4, the closed system polarizations are close to
1, and the 4 zone closed system polarizations can be increased to
the level of the 1 and 2 zone closed system polarizations by doubling
the adiabatic switching time to $40\hbar/E_{k}$. Note that at $T=300$
K, the polarizations of Cell 4 at the end of each process are significantly
lower than the closed system polarizations, indicating that the phonon
interaction has degraded reliability of information transmission.
This is because the kink energy $E_{k}=0.05$ eV for silicon dangling
bond circuits is close to the room temperature thermal energy $k_{B}T=0.026$
eV, so the total induced transition rate 

\begin{center}
\includegraphics[scale=1.0]{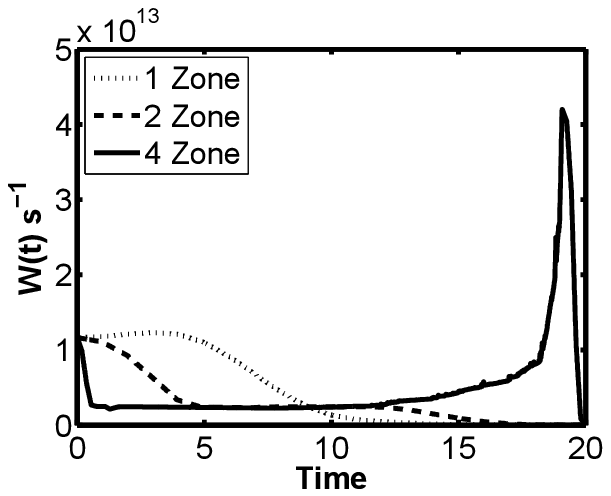}
\par\end{center}

\begin{flushleft}
{\small{}Fig. 5 Ground to first excited state transition rates as
a function of time throughout the duration of different clocking processes.}
\par\end{flushleft}{\small \par}

\noindent $W=W_{10}+W_{20}+W_{30}+W_{40}$ for each zoning is large
enough to have a significant effect over the process duration. Specifically,
the rates for each zoning are plotted in Fig. 5, indicating $max\thinspace W>10^{13}$
s\textsuperscript{-1}in each case, from which it follows that the
timescale for excitation out of the ground state $1/W<10^{-13}$ s
is short compared with the total adiabatic switching time to the duration
of adiabatic switching. In this regime, the Born approximation does
not hold, so the Lindblad equation is not strictly accurate, but the
results suggest the operating temperature may have to be lowered below
room temperature for short silicon dangling bond wires to function
effectively.

\noindent Lowering to liquid nitrogen temperature ($77$ K), the new
expected output cell polarizations after each process has been completed
are shown in Fig. 6.

\begin{center}
\includegraphics[scale=1.0]{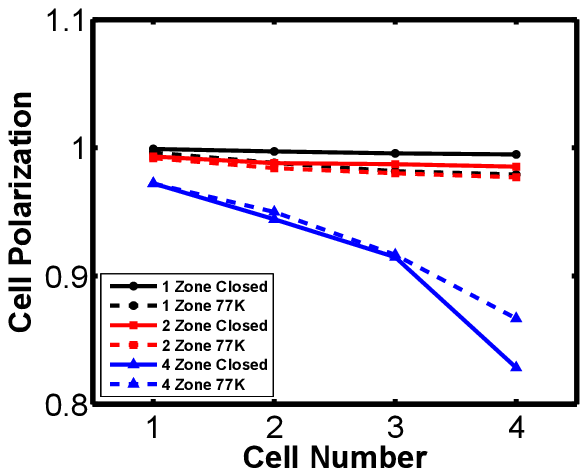}
\par\end{center}

\begin{flushleft}
{\small{}Fig. 6 Cell polarizations of four cell wire at the completion
of different clocking processes run at $T=300$ K.}
\par\end{flushleft}{\small \par}

\noindent The plots show that lowering the temperature has succeeded
in restoring wire performance to its closed system performance by
lowering the ground-to-excited state transition rates. Note that the
excited-to-ground state transition rates are still large, and this
is why the 4 zone cell 4 polarization actually increases when environmental
interactions are taken into account instead of decreasing like in
the other cases. Changing wire length, temperature, and QCA parameters,
it is not strictly true that using more zones protects wire performance
better against phonon interaction. However, when this is the case,
it is typically because many of the transition rates $W_{rs}$ vanish
during portions of the adiabatic switching cycle when multiple clocking
zones are used due to additional symmetry of the time dependent Hamiltonian
that occurs once a particular cell is latched. Specifically, from
the form of the wire Hamiltonian:

\begin{eqnarray}
H_{wire}(t) & = & -P_{in}(t)\frac{E_{k}}{2}\sigma_{z}^{1}+\frac{E_{k}}{2}\sum_{i=1}^{N}\gamma^{i}(t)\sigma_{x}^{i}\\
 &  & -\frac{E_{k}}{2}\sum_{i=1}^{N-1}\sigma_{z}^{i}\sigma_{z}^{i+1},\nonumber 
\end{eqnarray}

\noindent it follows that if $\gamma^{i}(t)=0$, then $\sigma_{z}^{i}$
commutes with $H_{wire}(t)$:

\begin{equation}
[\sigma_{z}^{i},H_{wire}(t)]=0.
\end{equation}

\noindent This means the Hamiltonian and $i^{th}$ spin can be simultaneously
diagonalized, or equivalently, that the energy eigenstates of $H_{wire}(t)$
may be split into two groups depending on the eigenvalue of $\sigma_{z}$.
In turn, this implies that if $|s_{+1}\rangle$ and $|s_{-1}\rangle$
are any two eigenstates of the Hamiltonian with different $i^{th}$
cell spin eigenvalues:

\begin{equation}
\sigma_{z}^{i}|s_{+1}\rangle=|s_{+1}\rangle,
\end{equation}

\begin{equation}
\sigma_{z}^{i}|s_{-1}\rangle=-|s_{-1}\rangle,
\end{equation}

\noindent then the induced phonon transitions between these states
is zero:

\begin{equation}
\langle s_{-1}|\sigma_{z}^{i}|s_{+1}\rangle=0.
\end{equation}

\noindent Furthermore, for any other operator $S$ that commutes with
$\sigma_{z}^{i}$:

\begin{equation}
[\sigma_{z}^{i},S]=0,
\end{equation}

\noindent we have:

\begin{equation}
\langle s_{-1}|[\sigma_{z}^{i},S]|s_{+1}\rangle=0,
\end{equation}

\begin{equation}
\Rightarrow\langle s_{-1}|\sigma_{z}^{i}S-S\sigma_{z}^{i}|s_{+1}\rangle=0,
\end{equation}

\begin{equation}
\Rightarrow\langle s_{-1}|-S-S|s_{+1}\rangle=0,
\end{equation}

\begin{equation}
\Rightarrow\langle s_{-1}|S|s_{+1}\rangle=0.
\end{equation}

\noindent Noting that $[\sigma_{z}^{i},\sigma_{z}^{j}]=[\sigma_{x}^{i},\sigma_{x}^{j}]=0$
for all cell indices $i$ and $j$, and that $[\sigma_{x}^{i},\sigma_{z}^{j}]=0$
for all indices $i\neq j$, it follows the operator $\beta(\vec{q})$
(equation (19)) which is a linear sum of these operators and responsible
for environementaly induced transitions (equations (16) and (17))
does not induce transitions between several pairs of states once multiple
cells are latched. In many cases this enhanced symmetry results in
improved wire performance in the presence of phonon interactions. 

\section{Gain and Dissipation}

\noindent The results of the previous section showed that if a short
4 cell silicon dangling bond wire begins in its ground state, then
room temperature phonon interactions induce transitions in the wire
at rates that make information transmission unreliable. However, lowering
to liquid nitrogen temperature restored successful information transmission
by increasing the timescale for excitation out of the ground state
to a value greater than the adiabatic switching interval. Determination
of whether or not reliable information transmission over long silicon
dangling bond wires initialized in their ground state is achievable
at low temperature requires different modeling techniques, since the
Born approximation of weak environmental interaction required to validate
use of the Lindblad equation no longer holds for sufficiently large
numbers of cells (see Appendix A).

\noindent Another issue is that even if we ignore environmentally
induced transitions that may occur during information transmission,
thermal degradation can also occur due to thermalization of the initial
wire state. That is, if the wire has been unused for a sufficiently
long duration, we expect it reside in a mixed state of thermal equilibrium
rather than its ground state. To investigate this, we neglect all
induced transitions ($W_{rs}=0$) and consider closed system processes
beginning from a state of thermal equilibrium. Fig. 7 shows the results
of simulating an information transmission process along an 8 cell
wire with fixed input polarization and all cells relaxed, starting
from the ground state ($T=0$ K). It plots the output polarizations
of each of the wire cells after 4 processes with different zonings
are completed. The total duration of each process is $40\hbar/E_{k}$.

\begin{center}
\includegraphics[scale=1.0]{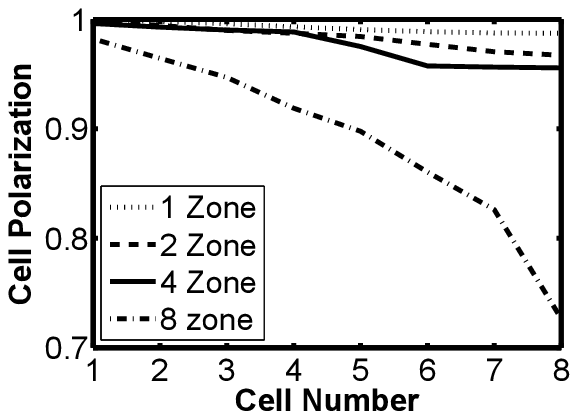}
\par\end{center}

\begin{flushleft}
{\small{}Fig. 7 Output cell polarizations of an eight cell wire after
different clocking processes beginning from thermal equilibrium.}
\par\end{flushleft}{\small \par}

\noindent The plot shows that when the wire is closed and 1,2, or
4 zones are used, the switching time is sufficiently long for high
fidelity information transmission. When 8 zones are used, the switching
of individual cells is faster and non-adiabatic transitions to excited
states occur and signifcantly increase the probability of error {[}12{]}. 

\noindent Fig. 8 shows two plots of the same information transmission
process beginning from thermal equilibrium states at $T=77$ K and
$T=300$ K, respectively. These plots indicate that regardless of
the zoning scheme, beginning transmission from a thermal state results
in polarization decay of the cells along the wire. The exponential
nature of the decay in polarization as described in Section I is clearly
evident at $T=300$ K where thermal errors dominate any transmission
errors due to non-adiabatic transitions when 8 clocking zones are
used.

\begin{center}
\includegraphics[scale=1.0]{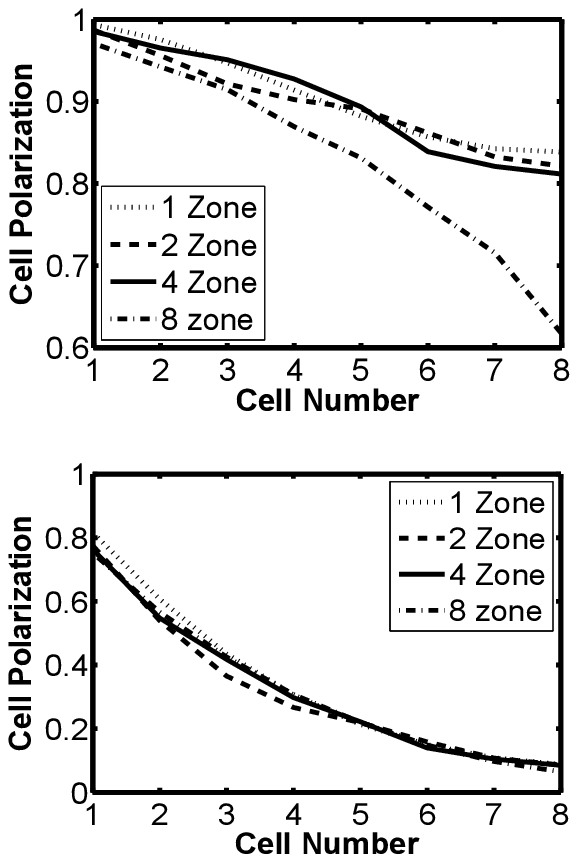}
\par\end{center}

\begin{flushleft}
{\small{}Fig. 8 Output cell polarizations along an 8 cell wire beginning
information transmission from thermal equilibrium at $T=77$ K (top)
and $T=300$ K (bottom).}
\par\end{flushleft}{\small \par}

\noindent We can also examine a different process with fixed input
polarization 1 which begins from a thermal equilibrium state at $T=300$
K in which all the cells are latched (clock signals are zero). In
this initial state, the cell polarizations decay exponentially along
the wire away from the input as in the introductory example of Section
I. From this state, each cell in the wire is relaxed and relatched
sequentially to transmit the input polarization along the wire. Fig.
9 shows the cell polarizations before and after the process. Once
again, the cell polarizations decay along the wire exponentially after
switching is finished.

\begin{center}
\includegraphics[scale=1.0]{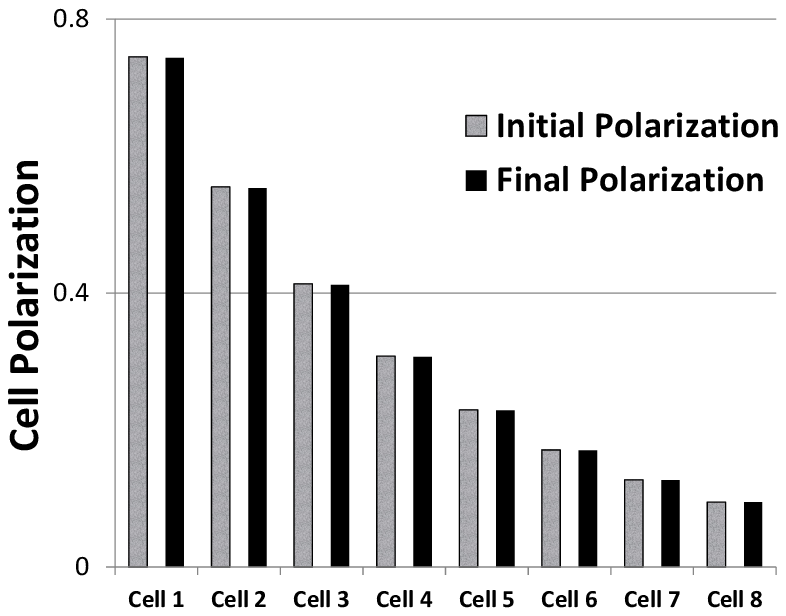}
\par\end{center}

\begin{flushleft}
{\small{}Fig. 9 Cell polarizations before and after a switching process
run from a latched thermal equilibrium state.}
\par\end{flushleft}{\small \par}

\noindent These results show that even at liquid nitrogen temperature,
operating a relatively short silicon dangling bond wire from thermal
equilibrium will result in exponential decay of polarization and limit
its ability to transmit information. Furthermore, the results hold
true regardless of the zoning scheme, so in this circumstance polarization
amplification is absent. That is, the expected output polarization
of a zone is not amplified, as measured by the expected polarization
of the next zone, so there is no gain in the polarization as we might
hope for. However, this is what we should expect for an adiabatic
process which evolves excited states into other excited states, since
this implies a thermally mixed state will evolve under coherent evolution
into a different mixture of excited states. It follows that if we
are to obtain the correct output polarization across a QCA wire operated
from thermal equilibrium, it must be a consequence of dissipative
environmental effects which can induce transitions from excited to
lower energy wire states during adiabatic switching. One possibility
is that induced transitions occur in such a manner that individual
zones are brought to thermal equilibrium quickly relative to their
adiabatic switching time. In this event, if the zones are sufficiently
small, each zone will be brought into its ground state with its cells
following the input polarizations with high probability. 

\noindent To investigate this possibility, for simplicity, we can
assume that zones consist of single cells. In this event, assuming
a particular cell in the wire experiences an input polarization $P\in[0,1]$,
then:

\begin{equation}
P=q(1)+(1-q)(-1)=2q-1,
\end{equation}

\noindent where $q$ is the probability that the input cell brought
to thermal equilibrium has polarization $+1$. Letting $p=e^{-E_{k}/k_{B}T}$,
it follows that the expected cell output polarization after thermal
equilibrium is reached is:

\begin{equation}
Pol=\frac{q+(1-q)p-(1-q)-qp}{q+(1-q)p+(1-q)+qp}=(2q-1)\frac{1-p}{1+p},
\end{equation}

\noindent which is smaller than the input polarization by a factor
of $(1-p)/(1+p)$. This implies polarization decay also occurs in
this situation. This is an important point, since in a model of the
cell where the input polarization is regarded as a continous function
{[}13{]} rather than a cell with discrete polarizations we would have:

\begin{equation}
Pol=\frac{1-e^{-P\cdot E_{k}/k_{B}T}}{1+e^{-P\cdot E_{k}/k_{B}T}},
\end{equation}

\noindent instead, and this model does exhibit polarization gain for
small input polarizations $P$ whenever $\frac{E_{k}}{k_{B}T}>2$:

\begin{equation}
Pol=\frac{1-e^{-P\cdot E_{k}/k_{B}T}}{1+e^{-P\cdot E_{k}/k_{B}T}}>P,
\end{equation}

\noindent It is easy to verify that the former polarization decay
formula holds instead of the polarization gain formula for a 2 cell
wire latching process with each cell clocked individually. Assuming
a kink energy of $0.05$ eV and operating temperature satisfying $k_{B}T=0.017$
eV, we can use the Lindblad equation with transition rates:

\begin{eqnarray}
W_{rs} & = & C\cdot|\langle\psi_{r}|\sigma_{x}^{1}|\psi_{s}\rangle|^{2},\\
W_{sr} & = & e^{(E_{s}-E_{r})/k_{B}T}W_{rs},
\end{eqnarray}

\noindent while the first cell is latched, and transition rates:

\begin{eqnarray}
W_{rs} & = & C\cdot|\langle\psi_{r}|\sigma_{x}^{2}|\psi_{s}\rangle|^{2},\\
W_{sr} & = & e^{(E_{s}-E_{r})/k_{B}T}W_{rs},
\end{eqnarray}

\noindent while the second cell is latched. These transitions act
to flip the \textit{z}-polarization of the first or second cell respectively,
and choosing the constant $C$ sufficiently large brings the first
and second cells into thermal equilibrium during their respective
latchings. At the end of this process, the simulation gives cell polarizations:

\begin{eqnarray}
Cell\thinspace1\thinspace Pol & = & 0.8997=\frac{1-p}{1+p},\\
Cell\thinspace2\thinspace Pol & = & 0.8095=\left(\frac{1-p}{1+p}\right)^{2}.
\end{eqnarray}

\noindent Note that the second polarization is not:

\begin{equation}
\frac{1-e^{-0.8997\cdot E_{k}/k_{B}T}}{1+e^{-0.8997\cdot E_{k}/k_{B}T}}=0.8675,
\end{equation}

\noindent as would be the case if the formula exhibiting polarization
gain held true. 

\section{Quantum Annealing}

\noindent Given that operating a QCA wire from thermal equilibrium
results in exponential polarization decay in both cases of purely
coherent evolution and strong decoherence, the question arises as
to whether or not there are any other techniques for bringing wires
into their ground state with high probability either before or during
information transmission. One way of doing this is to adiabatically
switch the wire through a state where at thermal equilibrium it occupies
its ground state with very high probability. For instance, in the
event that the tunneling coefficients $\gamma^{i}$ can be raised
to very large ($\gg E_{k}$) values, the wire cells may effectively
decouple, and the entire wire will reach thermal equilibrium as each
of its cells reaches its own state of thermal equilibrium. Therefore,
adiabatically switching the wire through this high tunneling regime
might be used as a means of restoring the wire ground state with high
probability at regular intervals to avoid occupying information scrambling
thermal states. This process is known as quantum annealing {[}14{]}.

\noindent For large $\gamma^{i}$, each cell will occupy its ground
state with high probability once it has equilibrated, since at equilibrium
this probability is:

\begin{equation}
p_{gnd}=\frac{1}{1+e^{-\gamma_{max}/k_{B}T}}.
\end{equation}

\noindent It follows that the probability of an $N$ cell wire occupying
its ground state is $p_{gnd}^{N}$. For 1000 completely decoupled
silicon dangling bond cells operating at liquid nitrogen temperature
and maximum tunneling $\gamma_{max}=0.05$ eV, the probability of
the array occupying its ground state after relaxation is 0.32, but
increases to 0.9993 when $\gamma_{max}=0.1$ eV. In practice, larger
maximum tunnelings may be required due to the\textit{ z}-polarization
coupling terms in the QCA Hamiltonian. If tunneling modulation can
be implemented in silicon dangling bond technology, it is possible
such maximum tunnelings are achievable, although a mechanism other
than phonon emission/absorption will be necessary to relax cells since
the longitudinal acoustic phonon spectrum only extends up to $0.063$
eV. Therefore, utilizing this strategy for silicon dangling bond QCA
requires further progress in engineering dangling bond cells with
large and tuneable tunneling and identifying environmental interactions
that can relax cells on a suitable timescale. 

\section{Conclusions}

\noindent Simulations of 4 cell silicon dangling bond wires suggest
they can operate reliably from their ground state at liquid nitrogen
temperature due to reduced environmentally induced excitations of
the ground state. However, simulations of longer wires initialized
in states of thermal equilibrium show exponential decay of cell polarization
along the wire regardless of clock zoning. Further theoretical arguments
and simulation suggest this polarization decay persists even when
environmental transitions bring wire zones to local thermal equilibrium
quickly. To solve this problem, a quantum annealing process whereby
cell tunnelings are ramped to a level significantly greater than the
intracell coupling is proposed as a method of periodically re-initializing
the wire in its ground state with high probability. In principle,
the same technique could be used to bring arbirary QCA circuits into
their ground state, although for realization this approach requires
engineering cells with large and tunable maximum tunneling, which
is an ongoing topic of research in the development of silicon dangling
bond technology. 

\section*{Acknowlegment}

\noindent This post-doctorate work at the University of British Columbia was supported by an NSERC Engage grant in parternship with Quantum Silicon Inc.

\section*{Appendix A: Validity of Master Equation}

\noindent The Lindblad equation introduced in Ssection 3:

\begin{eqnarray}
\frac{d}{dt}\rho_{wire} & = & -\frac{i}{\hslash}[H_{wire},\rho_{wire}]\\
 &  & +\sum_{r,s}\frac{1}{2}W_{rs}\left(2P_{rs}\rho_{wire}P_{sr}-[P_{sr}P_{rs},\rho_{wire}]\right),\nonumber 
\end{eqnarray}

\noindent is used to describe the interaction of a QCA wire with phonons
in its environment. The derivation of this equation from the coherent
quantum dynamics of the wire and phonon bath taken together relies
on using the Born, Markov, and secular approximations in succession.
Together, the correctness of the Born and Markov approximations imply
the applicability of the Bloch-Redfield master equation. The secular
approximation is then used to neglect particular terms in the Bloch-Redfield
master equation, giving rise to the Lindblad master equation. These
approximations are detailed below:
\begin{enumerate}
\item Born approximation: The effect of the interaction of the phonon bath
with the QCA wire is weak, and may be treated as a perturbation of
the coherent wire dynamics. This implies the sum of the induced transition
rates $W_{rs}$ into or out of any state is small compared to the
characteristic energy scale of the system and adiabatic clocking frequency:
\begin{equation}
\sum_{r}W_{rs}\ll E_{k}/\hbar=1.52\cdot10^{13}s^{-1}.
\end{equation}
\item Markov approximation: The fluctuations away from thermal equilibrium
of the phonon bath occur on a timescale that is much shorter than
the characteristic timescales of interaction of the phonon bath and
QCA wire as determined by the phonon induced transition rates:
\begin{equation}
\sum_{r}W_{rs}\ll\omega_{q_{max}}=1.06\times10^{14}s^{-1}.
\end{equation}
\item Secular approximation: All internal transition frequencies of the
QCA wire $\omega_{ij}=\hbar E_{ij}$ are significantly larger than
the phonon induced transition rates:
\begin{equation}
\sum_{r}W_{rs}\ll\hbar E_{ij},
\end{equation}
This condition implies various terms in the Bloch-Redfield master
equation average to zero. This is a more stringent condition than
the Born approximation, since any degeneracy of states in the QCA
wire will prevent this condition from holding true.
\end{enumerate}
\noindent Notably, the secular approximation does not generally hold
for adiabatic clockings of QCA circuit with multiple cells because
of degeneracies and near degeneracies of wire states. Because of this,
when the Born and Markov approximations hold, it is more accurate
to consider the QCA wire as satisfying a Bloch-Redfield master equation,
which written in the interaction picture is {[}8{]}: 

\begin{equation}
\dot{\rho}_{s's}^{(i)}(t)=\sum_{m,n}\gamma_{s'smn}(t)\rho_{mn}^{(i)}(t),
\end{equation}

\begin{eqnarray}
\gamma_{s'smn}(t) & = & [-\sum_{k}\delta_{sn}\Gamma_{s'kkm}^{+}+\Gamma_{nss'm}^{+}\\
 &  & +\Gamma_{nss'm}^{-}-\sum_{k}\delta_{sn}\Gamma_{s'kkm}^{+}]e^{i(\omega_{s's}-\omega_{mn})}.\nonumber 
\end{eqnarray}

\noindent In this equation, the $\Gamma^{\pm}$ terms are various
relaxation rates between states expressible in terms of the system-bath
interaction. The secular approximation assumes that only transition
rates $\gamma_{ssss}$, $\gamma_{ssmm}$ ($s\neq m$), and $\gamma_{s'ss's}$
are nonzero. In this case:

\begin{equation}
\gamma_{ssmm}(t)=W_{sm},
\end{equation}

\noindent This rough approximation is assumed in Section 4 for four
cell wires with only 16 total states, since the states for a single
clocking zone are typically non-degenerate. Also, many of the transition
rates that are ignored in making the secular approximation vanish
for much of adiabatic switching when multiple clocking zones and state
degeneracies occur. 

\noindent \smallskip

\noindent Having justified the use of the secular approximation for
short wires, we can now go back and analyze when we expect the Born
and Markov approximations to hold true, since we can expect that as
a QCA circuit gets larger, phonon induced transition rates will in
some sense begin to dominate its dynamics. Since the induced transition
rates vary during adiabatic switching, we can, as a rough estimate,
consider sums of induced rates $\sum W_{r0}$ at $T=77$ K out of
the ground state of $N$ cell silicon dangling bond arrays with zero
intercell coupling and input polarization, and tunneling energy $\gamma^{i}(t)=0.05$
eV. In this example, a phonon can be absorbed by any cell to excite
the wire ground state to the $N-$fold degenerate first excited state,
and the sum of transition rates is empirically:

\[
\sum W_{r0}\approx1.2\cdot10^{10}N,
\]

\noindent which is equal to the characteristic frequency $E_{k}/\hbar=1.52\cdot10^{13}$
s\textsuperscript{-1} when $N=1270$. Therefore, as an initial estimate
we can guess the Born approximation begins to fail at $T=77$ K for
wires approximately 1000 cells in length.

\section{References}

\noindent {[}1{]} C. Lent, P. Tougdaw, ``A device architecture for
computing with quantum dots'', \textit{Proc. IEEE,} vol. 85, no.
4, pp. 541 - 557, 1997.

\noindent {[}2{]} C. Lent, P. Tougaw, \textquotedblleft Bistable saturation
due to single electron charging in rings of tunnel junctions\textquotedblright ,
\textit{Journal of Applied Phys}ics, vol. 75, no. 8, pp. 4077 - 4080,
1994.

\noindent {[}3{]} C. Lent, B. Isaksen, and M. Lieberman, \char`\"{}Molecular
quantum-dot cellular automata\char`\"{}, \textit{Journal of the American
Chemical Society,} vol. 125, no. 4, pp. 1056 - 1063, 2003.

\noindent {[}4{]} L. Livadaru, P. Xue, Z. Shaterzadeh-Yazdi, G. DiLabio,
J. Mutus, J. Pitters, B. Sanders, R. Wolkow, ``Dangling-bond charge
qubit on a silicon surface'', \textit{New Journal of Physics}, vol.
12, no. 8, 083018, 2010.

\noindent {[}5{]} T. Kato, \char`\"{}On the adiabatic theorem of quantum
mechanics\char`\"{}, \textit{Journal of the Physical Society of Japan,}
vol. 5, no. 6, pp. 435 - 439, 1950.

\noindent {[}6{]} C. Lent, D. Tougaw, W. Porod, \char`\"{}Quantum
cellular automata: the physics of computing with arrays of quantum
dot molecules.\char`\"{} \textit{IEEE Workshop on Physics and Computation},
1994.

\noindent {[}7{]} G. Mahan, \textit{Many-Particle Physics, }Springer
Science and Business Media New York, 2000.

\noindent {[}8{]} G. Mahler, and A. Volker, \textit{\small{}Quantum
networks, Dynamics of open nanostructures}, Springer-Verlag Berlin
Heidelberg, 1998. 

\noindent {[}9{]} U. Bockelmann, G. Bastard, \char`\"{}Phonon scattering
and energy relaxation in two-, one-, and zero-dimensional electron
gases\char`\"{}, \textit{Physical Review B,} vol. 42, no. 14, 8947,
1990.

\noindent {[}10{]} S. Barrett, G. Milburn, \char`\"{}Measuring the
decoherence rate in a semiconductor charge qubit\char`\"{}, \textit{Physical
Review B,} vol. 68, no. 15, 155307, 2003.

\noindent {[}11{]} S. Faleev, F. Léonard. \char`\"{}Theory of enhancement
of thermoelectric properties of materials with nanoinclusions.\char`\"{}
arXiv preprint arXiv:0807.0260 (2008).

\noindent {[}12{]} C. Zener, \char`\"{}Non-adiabatic crossing of energy
levels.\char`\"{} \textit{Proceedings of the Royal Society of London
A: Mathematical, Physical and Engineering Sciences}, vol. 137, no.
833, pp. 696 - 702, 1932. 

\noindent {[}13{]} J. Timler, C. Lent, ``Power gain and dissipation
in quantum-dot cellular automata'', \textit{Journal of Applied Physics},
vol. 91, no. 2, pp. 823 - 831, 2002.

\noindent {[}14{]} G. Santoro, et al. \char`\"{}Theory of quantum
annealing of an Ising spin glass\char`\"{}, \textit{Science,} vol.
295, no. 5564 pp. 2427 - 2430, 2002.

\end{document}